\documentclass[aps,prl,a4print,reprint,amsmath,amssymb,showpacs]{revtex4-1}
\usepackage{bm}
\usepackage[dvips]{graphicx}

\begin{document}

\title{Direct measurement of single soft lipid nano-tubes: nano-scale information extracted in non-invasive manner}
\author{Akihisa Yamamoto}
\author{Masatoshi Ichikawa}
\email{ichi@scphys.kyoto-u.ac.jp}
\affiliation{Department of Physics, Kyoto University, Sakyo, Kyoto, 606-8502, Japan}
\date{\today}

\begin{abstract}
We investigated the dynamics of single soft nano-tubes of phospholipids to extract nano-scale information such as size of tube, which are several tens to hundreds of nano-meters thick. The scaling law of the nano-tube dynamics in quasi-2-dimensional space was revealed to be constituent with that of a polymer. The dynamic properties of the tubes obtained from direct observation by fluorescent microscopy, such as their persistence length, enable us to access the nano-scale characteristics through a simple elastic model of the membrane. The present methodology should be applicable to the nano-sized membrane structure in living cells.
\end{abstract}


\maketitle

Within organisms, the phospholipid membrane is a primitive compartment wall of small structures that can exhibit various topologies and morphologies, such as sphere, tube, etc. While there are too many examples to enumerate, several characteristic morphologies have been reported in endoplasmic reticulum~\cite{sparkes_ER}, neurons and mitochondria~\cite{lea_mitochondria}. Although the shapes of the membrane as observed in electron micrograph images are fixed or frozen, they must be dynamic in living cells. For example, the deformation of mitochondria has been observed depending on changes in the environment~\cite{hahn_mitochondria}. Such dynamic morphologies under non-equilibrium and reactive conditions could have biological significance and should be examined by physico-chemical methods. This is one of the motivations for identifing and measuring the dynamics of the morphology of nanoscopic membrane structures such as tubes.

Recent biophysical approaches that use model cell systems have been shown to be useful for examining the above problems. For instance, giant vesicles which exhibit several tens of micrometers in size have been used as a model system for studying their morphological transformation~\cite{hotani_liposome}, microdomain formation or lateral phase separation~\cite{veatch_phase}. The encapsuplation of biopolymers has been applied for mimicing biochemical functions~\cite{walde_vesicle, tsumoto_transcription, nomura_expression} and structures~\cite{hase_actin, negishi_polyelectrolytes}. Toxicity by an enzyme on a membrane has been observed as a deformation of the vesicle~\cite{wick_microinjection}, and affinity of the DNA on a membrane has been identified with transcriptional processes~\cite{kato_DNA, tsuji_DNA}. Complex system or membrane containing biomolecules have been realized in supported membrane~\cite{tanaka_supported}.

In the present paper, we report a novel observation method for identifying the dynamics of single soft nano-tubes, and suggest a new strategy for measuring their physical properties or nanoscopic structures in a noninvasive manner. Dynamical properties, including the diffusion constant, relaxation time and persistence length, are obtained directly through the use of fluorescent microscopy. Comparison of a model of membrane elasticity based on the Helfrich scheme~\cite{helfrich} to the statics of a fluctuated tube enables us to determine the thickness of individual nano-tubes. This strategy should be applicable to non-immersive \textit{in vivo} measurements within cellular organisms.

\begin{figure}[!bp]
	\centering
	\includegraphics[scale=1.0]{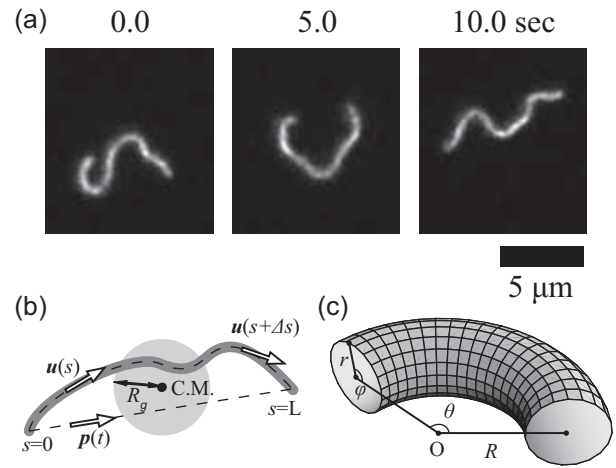}
	\caption{(a) Typical images of a single phospholipid tube in an aqueous solution. The images were taken at intervals of 5.0 s. The scale bar is 5 $\mathrm{\mu m}$. (b) A schematic illustration of a tube and the measured values obtained from image analysis. The radius of gyration $R_g$ is calculated from the spatial distribution of fluorescent intensity; the rotational relaxation time $\tau_r$ represents the decay of the correlation of a unit end-to-end vector $\bm{p}(t)$; the persistence length $l_p$ is defined as the decaying slope of a scalar product of two unit tangent vectors $\bm{u}(s)$ and $\bm{u}(s+\varDelta s)$. (c) Frame representation of part of a tube. An Eel-like tube is assumed to be composed of several sections of tori with various values of $R$ and homogeneous $r$. Notations used in the calculations are represented as above.}
	\label{snaps}
\end{figure}

Phospholipid tubes were made by the natural swelling method \cite{lasic_mechanism}. DOPC (Dioleoyl-L-$\mathrm{\alpha}$-phosphatidylcholine, purchased from Wako) and Rhodamine-Red DPPE (N-(rhodamine red-X)-1,2-dipalmitoyl L-$\mathrm{\alpha}$-phosphatidylethanolamine, Avanti Polar Lipids) at a molar ratio of $500:1$ were dissolved in chloroform/methanol ($ = 2:1$) solution, and then dried under vacuum overnight to form thin lipid films.
The dry films were hydrated with 200 $\mathrm{\mu L}$ of distilled water at room temperature for 3 hours. A glass chamber for microscopy observation was made using glass cover slips washed with alkaline alcohol. The swelled solution was placed between the slips so that the solution was 2 $\mathrm{\mu m}$ thick, and the chamber was sealed with hydrophobic liquid blocker to prevent evaporation. The sample chamber was set on a fluorescent microscope (Nikon TE2000U). Since the sample was confined to within this thin space that was on the order of the focal depth, the shape and motion of the whole tube could be observed as in-focus images. Sequential images were captured by an EM-CCD (Hamamatsu Photonics) and recorded at 31.93 frames/sec.

Figure \ref{snaps}(a) shows typical images of the time development of a single phospholipid tube exhibiting translational and intrachain Brownian motion. As shown, we observed individual tubes that were longer than several micrometers, and measured their positions and shape. The center of mass $\bm{r}_\mathrm{C.M.}$ of the phospholipid nano-tube is calculated as $\bm{r}_\mathrm{C.M.}=\sum_{i} \bm{r}_i \cdot I(\bm{r}_i)/\sum_{i} I(\bm{r}_i)$, where $I(\bm{r}_i)$ is the fluorescent intensity of each pixel $\bm{r}_i$.

\begin{figure}[!tbp]
	\centering
	\includegraphics[scale=1.0]{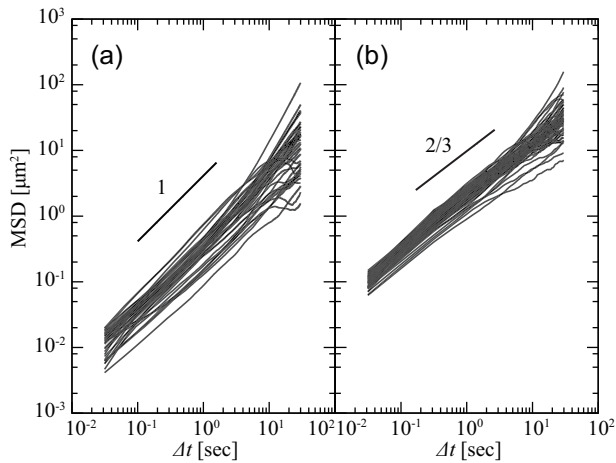}
	\caption{(a) Log-log plot of measurements of MSD of the center of mass. In the short time region to 10 sec, each MSD is proportional to $\varDelta t$. (b) Log-log plot of MSD of the end points.}
	\label{dif_log}
\end{figure}%

\begin{figure}[!tbp]
	\centering
	\includegraphics[scale=1.0]{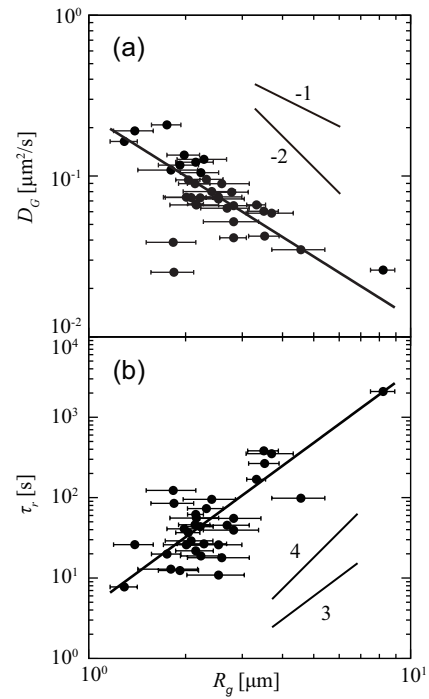}
	\caption{Relations between (a) the lateral diffusion constant $D_G$ and the radius of gyration $R_g$ and (b) the rotational relaxation time $\mathrm{\tau}_r$ and $R_g$ as determined experimentally. Fitting lines are $D_G \propto R_g^{\ \alpha} $ ($\alpha = -1.3$) and $\tau_r \propto R_g^{\ \beta}$ ($\beta = 2.9$), respectively. Superimposed lines show the scaling law from the theory of polymer dynamics; $\alpha = -2$ and $\beta = 4$ are indices of the Rouse model, and $\alpha = -1$ and $\beta = 3$ are those of the Zimm model.}
	\label{D-tau-Rg_d2um}
\end{figure}%

The mean square displacement (MSD) of $\bm{r}_{\mathrm{C.M.}}$ for each tube is shown in Fig. \ref{dif_log}(a) as a function of the lag time $ \varDelta t $.  From the slopes of MSD in the short time region less than 10 sec, the diffusion coefficient $D_G$ is calculated as $\mathrm{MSD} = 4D_G\varDelta t + A(\varDelta t)^2$, where $A$ is a drift coefficient.

Figure \ref{dif_log}(b) shows the MSD of the end points of each tube and the average of scaling indices in the short time region less than 10 sec, that is determined to be 0.76 $\pm$ 0.18. The lateral motion of an end point can be predicted by a simple scaling law that the index is calculated to be 2/3 before the configuration undergoes relaxation~\cite{doi_polymer}, and thus the experimental results show quite good agreement with the model of polymer dynamics.

As shown in Fig. \ref{snaps}(b), the radius of gyration and the unit end-to-end vector are also obtained as $R_g^2= \sum_{i}I(\bm{r}_i)\cdot |\bm{r}_i-\bm{r}_\mathrm{C.M.}|^2/\sum_{i} I(\bm{r}_i)$ and $\bm{p}(t)= (\bm{r}_{s=\mathrm{L}}(t) - \bm{r}_{s=0}(t))/\left| \bm{r}_{s=\mathrm{L}}(t) - \bm{r}_{s=0}(t) \right| $, respectively. $D_G$ for each tube is plotted as a function of $R_g$ in Fig. \ref{D-tau-Rg_d2um}(a). The scaling index $\alpha$ of $D_G \propto R_g^{\ \alpha}$ is $\alpha = -1.3 \pm 0.2$. The superimposed lines in Fig. \ref{D-tau-Rg_d2um}(a), $\alpha = -2 $ and $\alpha = -1$, indicate the indices calculated from the Rouse model and Zimm model for polymers, respectively~\cite{doi_polymer}. The rotational relaxation time $\tau_r$ of the unit of end-to-end vector $\bm{p}(t)$ is also obtained in Fig. \ref{D-tau-Rg_d2um}(b) as a function of $R_g$. The scaling index $\beta$ for $D_G \propto R_g^{\ \beta}$ in the experiment is estimated to be $\beta = 2.9 \pm 0.2$. The indices for the Rouse $\beta = 4$ and Zimm $\beta = 3$ models are also shown~\cite{doi_polymer}. The indices from the experiment are rather close to those of Zimm.

The experiments show that the quasi-two-dimensional chamber enables the observation of the dynamics of whole soft nano-tubes. Since, the yielded dynamics are similar or comparable to the previous studies ~\cite{barry_rodlike, kas_factin, goff_factin, han_ellipsoid}, thus the result itself must be reliable. In other words, the experimental results show that phospholipid nano-tubes behave as a polymer chain of the order of micrometers long.

Next, we discuss and figure out a simple methodology or strategy for determining the unseeable microscopic features of the membrane from observable values by coupling polymer physics and membrane physics through fluctuation-response.

\begin{figure}[!tbp]
	\centering
	\includegraphics[scale=1.0]{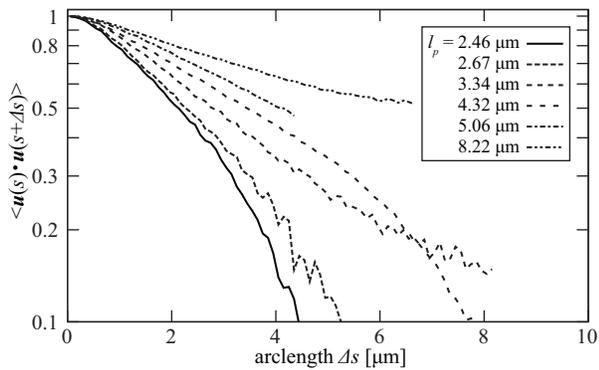}
	\caption{Typical examples of the correlations of scalar product $\langle \bm{u} (s) \cdot \bm{u}(s+\varDelta s) \rangle$. The persistence length $l_p$ is determined from the exponential decay in a short region. The correlation at long range and in the vicinity of 0 tends to deviate from exponential fitting.}
	\label{100602-002CCF}
\end{figure}%

Geometrically, it is easy to assume a bent tube to be comprised of parts of a torus as shown in Fig. \ref{snaps}(c). The bending free energy of the membrane $F$ can be denoted by a Helfrich-type expression as~\cite{helfrich},
\begin{equation}
	F=\frac{\kappa}{2}\int(C_1+C_2-C_0)^2 \mathrm{d}A.
\end{equation}
For part of a torus, $C_1$ and $C_2$ are curvatures that are normal to each other on the membrane surface. $C_0$ is the spontaneous curvature and in this case is assigned a value of zero. When the small radius and its angle are defined as $r$ and $\phi$ for the $C_1$ coordinate, and the large radius and its angle are $R$ and $\theta$ for the $C_2$ coordinate, $C_1 = \frac{1}{r}$ and $C_2 = \frac{\cos \phi}{R+r\cos \phi}$ are given. The area element $\mathrm{d}A$ is represented as $\mathrm{d}A = r\mathrm{d}\phi \cdot (R+r\cos \phi)\mathrm{d}\theta $. Thus, the bending energy of a torus for which the cross-sectional circle sweeps in small angle $\varDelta \theta$ becomes,
\begin{equation}
	\varDelta F_{\mathrm{torus}} = \frac{\kappa}{2} \frac{2\pi}{r}\frac{1}{\sqrt{1-r^2 c^2}}\varDelta s, \label{Ftorus0}
\end{equation}
where $c$ is the curvature of the circular arc drawn by the backbone of the torus, and $\varDelta s = R \varDelta \theta$ corresponds to the small section of the arc-length \cite{sup1}. Hence, the Taylor series of $\varDelta F_{\mathrm{torus}}$ for $c$ becomes,
\begin{equation}
	\varDelta F_{\mathrm{torus}}=\frac{\kappa \pi}{r} \varDelta s + \frac{r \kappa \pi}{2} \varDelta s c^2 + \frac{3r^3\kappa \pi}{8} \varDelta s c^4 + \cdots . \label{Ftorus}
\end{equation}
The first term on the right-hand side represents the bending energy for making a tube from a plane membrane, and the second and later terms are the result of bending of the backbone.

On the other hand, the bending free energy of a chain such as a polymer embedded in two spatial dimensions can be represented as~\cite{landau_elasticity, gutjahr_semiflexible},
\begin{equation}
	\varDelta F_{\mathrm{chain}} = \frac{k_\mathrm{B}Tl_{p}}{4} \varDelta s c^2, \label{Fpolymer}
\end{equation}
where $l_{p}$ is the persistence length of a chain in two spatial dimensions. If we compare the $c^2$ terms of eqs.~\eqref{Ftorus} and~\eqref{Fpolymer},
\begin{equation}
  \kappa = \frac{k_\mathrm{B}T l_p}{2 \pi r}, \label{kappa1}
\end{equation}
is obtained as a first approximation.

\begin{figure}[!tbp]
	\centering
	\includegraphics[scale=1.0]{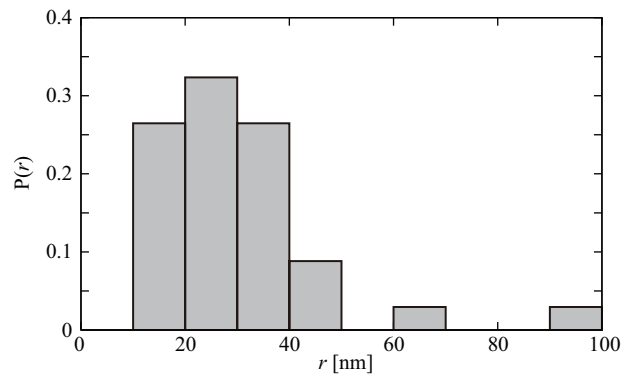}
	\caption{Distribution of the radii of tubes with $\kappa = 20 \ k_\mathrm{B} T$.}
	\label{radius}
\end{figure}

The persistence length $l_p$ of a nano-tube can be obtained from the experiments as
\begin{equation}
	\langle \bm{u}(s) \cdot \bm{u}(s + \varDelta s) \rangle = \exp \left( -\frac{\varDelta s}{l_p} \right), \label{lp}
\end{equation}
where $\bm{u}(s)$ is a two-dimensional unit tangent vector at which $s$ is the path length along the backbone from one end, as shown in Fig. \ref{snaps}(b). The value of $l_p$ for each tube can be estimated from the slope in Fig. \ref{100602-002CCF}. Short-range correlation lower than 0.5 $\mu$m is excluded from the fitting analysis~\cite{Ott_actin, dogic_nematic}. The above equation includes an approximation which adapts the present geometry and the order of the persistence length: $l_p$ calculated from the two-dimensional projection image of the polymer chain within a quasi-two-dimensional space asymptotically converges to the actual two-dimensional persistence length~\cite{hendricks_confined}. The present experimental values, thickness of the confining space and the order of magnitude of the persistence length, satisfy the above approximation.

The bending rigidity of a DOPC bilayer membrane has been experimentally measured by various methods, such as the pipette injection method~\cite{rawicz_effect, bo_elastic} and the electrodeformation method~\cite{gracia_electrodeformation}, and is known to be approximately 20 $k_\mathrm{B} T$. Therefore, the thickness of each nano-tube can be calculated from eq.~\eqref{kappa1} by substituting the value into $\kappa$, and the result is shown in Fig. \ref{radius}. The radii of nano-tubes, which are evaluated individually, are distributed around 30 nm. This value is consistent with the microscopy observation, considering the limitations due to resolution and diffraction limit, where the thickness is less than or equal to the fluorescent blurring length. In this manner, dynamical analysis of the micrometer-scale behaviors has revealed the size of a nano-structure.

The radii of the tubes can be estimated through another approach by considering the tube as a polymer chain divided into $N = L/b$ segments, where $L$ is the contour length and $b=2l_p$ is the Kuhn length. The diffusion constant of a rod with length $b$ and radius $r$ in two spatial dimensions, $D_S$, is calculated as~\cite{doi_polymer}
\begin{equation}
	D_S = \frac{3}{8} \frac{k_\mathrm{B}T\ln{(b/2r)}}{\pi \eta_s b}, \label{Ds}
\end{equation}
where $\eta_s$ is the viscosity of bulk water. $D_S$ can also be represented as $D_S = ND_G = L D_G/b$ in the Rouse model. Although the compensations for the entire length of the tube and the increase in viscosity from the walls are considerable \cite{lin_walls, supply}, a simple expression in terms of $r$ can be obtained as
\begin{equation}
	\frac{l_p}{r} = \exp \left( \frac{8\pi \eta_s}{3k_\mathrm{B} T}  L D_G \right). \label{b/r}
\end{equation}
Since only $L$ and $D_G$ are indefinite but experimentally observable variables, both $r$ and the bending modulus $\kappa$ of the membrane can be estimated from eqs.~\eqref{kappa1} and~\eqref{b/r}. However, the calculated values of $r$ obtained from eq.~\eqref{b/r} were scattered due to the exponential term, and the form is too sensitive to evaluate $r$ and $\kappa$ here.

The present discussion enables us to extract information regarding nanometer-scale properties of individual tubes from the dynamics, such as the size and bending modulus. Under optical microscopy, diffraction limitations or fluorescent blurring of approximately 200 nm in length hides nanoscopic structures. Based solely on a spatial analysis, the apparent sizes of the tubes in the fluorescent images were close to around 500 nm in diameter~\cite{grumelard}. On the other hand, the present method can access a scale that is one order of magnitude smaller than the length of optical resolution. This would be advantageous in measurements of living cells compared to methods that involve freezing or fixing, e.g., electron microscopy, which have thus far been used to measure nano-sized structures \cite{grumelard, imae_nanotubule}. Living cells and nonequilibrium conditions also require non-immersive approaches. Optical tweezers are useful tools for measurements in living cells compared to a tiny needle or micropipette~\cite{rawicz_effect, bo_elastic, ichikawa_optical, shitamichi_optical}, but they can be difficult to use with an internal structure confined in a cell. However, the present approach is free from such problems because of avoiding adhesive manipulations and modifications. Thus, this approach enables the observation of both living cells and non-equilibrium vesicles. For example, F-BAR protein families which are expected to induce budding or tube-generating transition in a living cell also make a spherical vesicle into tubules {\it in vitro} model experiment. By use of the present methodology, protein associated membrane tubes have been analyzed and yielded a mean bending modulus of the membrane reinforced by the proteins as a time-development behavior~\cite{takiguchi}. This non-immersive measurement would be a powerful method for evaluating nanoscopic sizes and properties in transient or nonequilibrium processes.

Although the model described here is constructed under the approximation of an elastic membrane, future work should consider the implications of the fact that a bilayer membrane is a liquid membrane. First, there is friction between the membrane and solution. This friction is discussed in plane geometry \cite{ramachandran_membrane}, but internal friction within the tube is more difficult to consider due to coupling of the shape of the tube vesicle. Another important consideration is thickness uniformity. The bent tube in this analysis is assumed to have a uniform thickness that does not depend on the backbone curvature $c$. However, the bending energy is also a function of the tube radius $r$, and not only $c$, as in the bending form of eq.~\eqref{Ftorus0} or~\eqref{Ftorus}. $r$ has a strong effect in higher-order terms. Nonlinear elastic behavior of this type is sometimes observed in experiments, and it would be interesting to quantify. These factors may affect those on a larger scale, i.e., those in eqs.~\eqref{lp} and~\eqref{Ds}. The nonlinearity of single chains might complement the rheology on a macroscopic scale, not only for tubes but also for a worm-like micelle solution \cite{mair_micelle_NMR}.

Zimm model takes into account hydrodynamic interaction by considering Oseen Tensor. The relative strength of the interaction depends on the thickness compared with the persistence length of the tube. Therefore, relative hydrodynamic interaction decays between thin segments in effectively swelled chains. This would be one of reasons of the behavior closing to that of Rouse model. Of course, hydrodynamic interaction through the inner part of the tube should be effective to the dynamics, but it is now an open question.

In conclusion, this is the first report on the use of direct observation to analyze the dynamics of single phospholipid soft nano-tubes. We have described a method for estimating the thickness and rigidity of a phospholipid nano-tube from the observed Brownian motion. The ability to avoid damaging the sample object should be useful not only for measuring the dynamical response of a membrane under non-equilibrium conditions but also for revealing the internal properties of membrane-closed organelles such as mitochondria and endoplasmic reticulum. 

We acknowledge Prof. K. Yoshikawa for his helpful discussion, and Dr. M. H\"orning and Dr. Y. Maeda for a critical reading of the manuscript. We also thank Dr. A. Isomura and Dr. T. Yamanaka for their help with analysis.
This work was supported by iCeMS Cross-Disciplinary Research Promotion Project and the grant-in-aid for the Grobal COE Program ``The Next Generation of Physics, Spun from Universality and Emergence'' from the Ministry of Education, Culture, Sports, Science and Technology (MEXT) of Japan.


\begin{thebibliography}{99}

\bibitem{sparkes_ER}I. A. Sparkes, L. Frigerio, N. Tolley and C. Hawes, Biochem. J. \textbf{423}, 145 (2009).

\bibitem{lea_mitochondria}P. J. Lea, R. J. Temkin, K. B. Freeman, G. A. Mitchell and B. H. Robinson, Microsc. Res. Tech. \textbf{27}, 269 (1994).

\bibitem{hahn_mitochondria}J. Bereiter-Hahn and M. V\"oth, Microsc. Res. Tech. \textbf{27}, 198 (1994).

\bibitem{hotani_liposome}H. Hotani, F. Nomura and Y. Suzuki, Curr. Opin. Colloid Interface Sci. \textbf{4}. 358 (1999).

\bibitem{veatch_phase}S. L. Veatch and S. L. Keller, Biochim. Biophys. Acta \textbf{1746}, 172 (2005).

\bibitem{walde_vesicle}P. Walde, K. Cosentino, H. Engel and P. Stano, ChemBioChem \textbf{11}, 848 (2010).

\bibitem{tsumoto_transcription}K. Tsumoto, S-i. M. Nomura, Y. Nakatani and K. Yoshikawa, Langmuir \textbf{17}, 7225 (2001).

\bibitem{nomura_expression}S-i. M. Nomura, K. Tsumoto, T. Hamada, K. Akiyoshi, Y. Nakatani and K. Yoshikawa, ChemBioChem \textbf{4}, 1172 (2003).

\bibitem{hase_actin}M. Hase and K. Yoshikawa, J. Chem. Phys. \textbf{124}, 104903 (2006).

\bibitem{negishi_polyelectrolytes}M. Negishi, M. Ichikawa, M. Nakajima, M. Kojima, T. Fukuda and K. Yoshikawa, Phys. Rev. E \textbf{83}, 061921 (2011).

\bibitem{wick_microinjection}R. Wick, M. I. Angelova, P. Walde and P. L. Luisi, Chem. Biol. \textbf{3}, 105 (1996).

\bibitem{kato_DNA}A. Kato, E. Shindo, T. Sakaue, A. Tsuji and K. Yoshikawa, Biophys. J. \textbf{97}, 1678 (2009).

\bibitem{tsuji_DNA}A. Tsuji and K. Yoshikawa, J. Am. Chem. Soc. \textbf{132}, 12464 (2010).

\bibitem{tanaka_supported}M. Tanaka and E. Sackmann, Nature \textbf{437}, 656 (2005).

\bibitem{helfrich}W. Helfrich, Z. Naturfors. \textbf{28c}, 693 (1973).

\bibitem{lasic_mechanism}D. D. Lasic, Biochem. J. \textbf{256}, 1 (1988).

\bibitem{doi_polymer}M. Doi and S. F. Edwards, \textit{The Theory of Polymer Dynamics} (Oxford Science Publications, 1988).

\bibitem{barry_rodlike}E. Barry and D. Beller and Z. Dogic, Soft Matter \textbf{5}, 2563 (2009).

\bibitem{kas_factin}J. K\"{a}s and H. Strey and J. X. Tang and D. Finger and R. Ezzell and E. Sackmann and P. A. Janmey, Biophys. J. \textbf{70}, 609 (1996).

\bibitem{goff_factin}L. L. Goff and O. Hallatschek and E. Frey and F. Amblard, Phys. Rev. Lett. \textbf{89}, 258101 (2002).

\bibitem{han_ellipsoid}Y. Han and A. M. Alsayed and M. Nobili and J. Zhang and T. C. Lubensky and A. G. Yodh, Science \textbf{314}, 626 (2012).

\bibitem{sup1}See Supplemental Material.

\bibitem{landau_elasticity}L. D. Landau and E. M. Lifshitz, \textit{Statistical Physics, 3rd Ed., Course of Theoretical Physics, Vol. 5} (Pergamon Press, 1994).

\bibitem{gutjahr_semiflexible}P. Gutjahr, R. Lipowsky and J. Kierfeld, Europhys. Lett. \textbf{76}, 994 (2006).

\bibitem{Ott_actin}A. Ott, M. Magnasco, A. Simon and A. Libchaber, Phys. Rev. E \textbf{48}, R1642 (1993).

\bibitem{dogic_nematic}Z. Dogic, J. Zhang, A. W. C. Lau, H. Aranda-Espinoza, P. Dalhaimer, D. E. Discher, P. A. Janmey, R. D. Kamien, T. C. Lubensky and A. G. Yodh, Phys. Rev. Lett. \textbf{92}, 125503 (2004).

\bibitem{hendricks_confined}J. Hendricks, T. Kawakatsu, K. Kawasaki and W. Zimmermann, Phys. Rev. E \textbf{51}, 2658 (1995).

\bibitem{rawicz_effect}W. Rawicz, K. C. Olbrich, T. McIntosh, D. Needham and E. Evans, Biophys. J. \textbf{79}, 328 (2000).

\bibitem{bo_elastic}L. Bo and R. E. Waugh, Biophys. J. \textbf{55}, 509 (1989).

\bibitem{gracia_electrodeformation}R. S. Gracia, N. Bezlyepkina, R. L. Knorr, R. Lipowsky and R. Dimova, Soft Matter \textbf{6}, 1472 (2010).

\bibitem{lin_walls}B. Lin, J. Yu and S. A. Rice, Phys. Rev. E \textbf{62}, 3909 (2000).

\bibitem{supply}See Supplemental Material.

\bibitem{grumelard}J. Grumelard, A. Taubert, W. Meier, Chem. Commun. \textbf{2004}, 1462 (2004).

\bibitem{imae_nanotubule}T. Imae, K. Funayama, M. P. Krafft, F. Giulieri, T. Tada and T. Matsumoto, J. Colloid Interf. Sci. \textbf{212}, 330 (1999).

\bibitem{ichikawa_optical}M. Ichikawa and K. Yoshikawa, Appl. Phys. Lett. \textbf{79}, 4598 (2001).

\bibitem{shitamichi_optical}Y. Shitamichi, M. Ichikawa and Y. Kimura, Chem. Phys. Lett. \textbf{479}, 274 (2009).

\bibitem{takiguchi}K. Takiguchi \textit{et al.}, in preparation.

\bibitem{ramachandran_membrane}S. Ramachandran and S. Komura, J. Phys.: Condens. Matter \textbf{23}, 072205 (2011).

\bibitem{mair_micelle_NMR}R. W. Mair and P. T. Callaghan, Europhys. Lett. \textbf{36}, 719 (1996).

\end{thebibliography}
\end{document}


\title{Supplemental Material for ``Direct measurement of single soft lipid nano-tubes: nano-scale information extracted in non-invasive manner''}
\author{Akihisa Yamamoto}
\author{Masatoshi Ichikawa}

\maketitle

\section{1. Derivation of free energy of membrane}

The non-invasive measurement for nano-meter sized objects is extensively studied in nanotechnology and condensed matter physics. Less explored but relevant one is membrane nano-tube that is widely seen in synthetic lipid vesicles and plasme membranes in living cells. In this studt we have developed novel method to infer the cross-section of fluctuating single membrane nano-tube. This supplemental material describes the brief summary for the calculation of the bending energy in eq. (1) shown in the manuscript. For general case, spontaneous curvature $C_0$ is regarded as non-zero variable.

As shown in the manuscript, two principal curvatures on the membrane surface are $C_1 = 1/r$ and $C_2 = \frac{\cos \phi}{R + \cos \phi}$, where $r$ and $\phi$ are the polar coordinates representing the cross-section of membrane nano-tube and $R$ is the radius of the backbone. Then, Eq. (1) of a torus which the cross-sectional circle sweeps in small angle $\varDelta \theta$ is written as
\begin{align}
  \varDelta F_\mathrm{torus} &= \frac{\kappa}{2} \int \left( \frac{1}{r} + \frac{\cos \phi}{R + \cos \phi} - C_0 \right)^2 r \left( R + r \cos \phi \right) \mathrm{d} \phi \mathrm{d} \theta \notag \\
&= \frac{\kappa}{2} \varDelta \theta \int_0^{2 \pi} \mathrm{d} \phi \left(b + \frac{\cos \phi}{a + \cos \phi} \right)^2 \left( a + \cos \phi \right), \tag{S1} \label{S1}
\end{align}
where $a=R/r > 1$ and $b=1-C_0 r$. We then rewrite Eq. \eqref{S1}:
\begin{equation}
	\frac{\varDelta F_\mathrm{torus}}{\kappa \varDelta \theta /2} = \int_0^{2 \pi}\mathrm{d} \phi \left( \left( b+1 \right) ^2 \cos \phi +ab^2-\frac{a \cos \phi}{a+\cos \phi} \right). \tag{S2} \label{S2}
\end{equation}
The integrals of first and second terms in Eq.~\eqref{S2} are $0$ and $2\pi a b^2$ respectively. With the complex integration on the unit circle $C$, we obtain the integral of third term as
\begin{align}
	&- a \oint_C \frac{(z+z^{-1})/2}{a+(z+z^{-1})/2} \cdot \frac{\mathrm{d}z}{iz} \notag \\
	&= - \frac{a}{i} \oint_C \frac{z^2+1}{z(z+a-\sqrt{a^2-1})(z+a+\sqrt{a^2-1})} \mathrm{d} z. \tag{S3} \label{S3}
\end{align}
This integrand has two single poles at $z_0=0$ and $z_1=-a+\sqrt{a^2-1}$ while we do not take into account $z_2=-a-\sqrt{a^2-1}$ is not the pole because it is out from the unit circle. The residues are given by
\begin{align}
 \mathrm{Res} \left( z_0 \right) & = \lim_{z \rightarrow z_0} \left \{ z \cdot \frac{z^2+1}{z(z-z_1)(z-z_2)} \right \} \notag \\
 & = 1 \tag{S4} \label{S4} \\
 \mathrm{Res} \left( z_1 \right) & = \lim_{z \rightarrow z_1} \left \{ (z-z_1) \cdot \frac{z^2+1}{z(z-z_1)(z-z_2)} \right \} \notag \\
 & = \frac{z_1^2+1}{z_1(z_1-z_2)} \notag \\
 & = -\frac{a}{\sqrt{a^2-1}}. \tag{S5} \label{S5}
\end{align}
We use the residue theorem and then rewrite Eq.~\eqref{S3} as
\begin{equation}
- \frac{a}{i} \cdot 2\pi i \sum_{j=0,1} \mathrm{Res} \left( z_j \right) = 2 \pi \left( \frac{a^2}{\sqrt{a^2-1}} -a \right) \tag{S6} \label{S6}
\end{equation}

Accordingly, we arrive at the following expression for eq.~\eqref{S1}:
\begin{align}
\varDelta F_\mathrm{torus} &= \frac{\kappa \varDelta \theta}{2} \cdot 2 \pi \left( \frac{a^2}{\sqrt{a^2-1}} -a +ab^2 \right) \notag \\
&= \frac{\kappa}{2} \frac{2 \pi}{r} \left( \frac{1}{\sqrt{1-r^2 c^2}} + C_0 r \left(C_0 r - 2 \right) \right) \varDelta s. \tag{S7} \label{S7}
\end{align}
If $C_0 =0$, one gets Eq. (2) in the manuscript. We note that $C_0 r \left(C_0 r - 2 \right)$ may have no effect in the bending of the backbone, so one can apply Eq. (5) written in the manuscript even though the spontaneous curvature of the membrane $C_0$ is non-zero. It just displaces zero point of the free energy.

\section{2. Configurational fluctuation in the direction of depth}

We next discuss additional effects of three-dimensional fluctuation of tube and viscosity compensation on the present geometry of membrane tube. A possible error in inferring the size of membrane tube may come from the fluctuation in the direction of depth because it makes the contour length $L_2$ that was observed in experiment shorter than the actual overall length $L$. For simplicity, a single wave approximation is applied in the present geometry as
\begin{equation}
z (s) =\frac{\varDelta z}{2} \sin(\frac{2\pi}{\lambda}s). \qquad (0\le s\le L_\mathrm{2}) \tag{S8} \label{S8}
\end{equation}
This displacement is introduced independent of bending in the $xy$-plane. The amplitude $\frac{\varDelta z}{2}$ and wavelength $\lambda$ of displacement are determined by the following considerations. Since the entire image of a tube is clearly seen during observation, the effective thickness is determined to be $\varDelta z=1\ \mathrm{\mu m}$. In terms of wavelength $\lambda$, in nature, persistence length $l_p$ may correspond to the combination of statistical fluctuation in the $xy$-plane and sine-wave displacement $z (s)$. For simplicity one may take a simple case with $\frac{\lambda}{2}=b$. From these considerations, $L$ can be represented as
\begin{equation}
	L = \int_0^{L_{2}} \mathrm{d}s \sqrt{1+\left( \frac{\varDelta z}{2} \frac{\pi}{2l_p}\right) ^2 \cos^2\left( {\frac{\pi}{2l_p}s}\right) }, \label{S8} \tag{S8}
\end{equation}
where $\varDelta z = 1 \ \mathrm{\mu m}$. In this expression, $L$ is 1.0 to 1.1 times longer than $L_2$ obtained from the projected image.

Another effect is the compensation of viscosity between two nearby walls. This can be described for a first approximation as 
\begin{equation}
	\eta _s (r) = \frac{\eta_{s0}}{1 - \frac{9}{16} ( \frac{r}{z} + \frac{r}{d-z} )} \label{S9} \tag{S9}
\end{equation}
, where $\eta_{s0}$, $d$ and $z$ are the viscosity of bulk water, the distance between two flat walls and the average distance to the central axis of the tube from the wall, respectively~\cite{lin_walls}. If we incorporate eq.~\eqref{S8} into $\eta_s$ of eq. (8) in the manuscript, the equation becomes nonlinear equation about $r$. One can get $r$ as well as $\eta_{s}(r)$ which becomes 1.0 to 1.1 times larger than $\eta_{s0}$.